\documentclass[journal,onecolumn]{IEEEtran}
\usepackage{setspace}
\usepackage{tabularx}
\usepackage[square]{natbib}
\usepackage{hyperref}

\usepackage{enumerate}

\ifCLASSINFOpdf
   \usepackage[pdftex]{graphicx}
\else
   \usepackage[dvips]{graphicx}
\fi
\usepackage[cmex10]{amsmath}
\usepackage{amsfonts}
\usepackage{amssymb}
\usepackage{amsthm}
\theoremstyle{plain}

\usepackage{booktabs}  
\usepackage{ragged2e} 
\newcolumntype{Y}{>{\RaggedRight\arraybackslash}X}

\DeclareMathOperator*{\argmax}{arg\,max}

\begin{document}
\doublespacing
%
\title{Distributed Resource Management in Systems of Systems: An Architecture Perspective \thanks{This is the pre-print version of the following article: \emph{M. Mosleh, P. ludlow, and B. Heydari, Distributed Resource Management in Systems of Systems: A Architecture Perspective, Systems Engineering, 2016}, which has been published in final form at 
doi: \href{http://doi.org/10.1002/sys.21342} {10.1002/sys.21342}      }}
%
%
%
\author{Mohsen~Mosleh,
        Peter~Ludlow,
        and~Babak~Heydari
\thanks{ Mohsen Mosleh, Peter Ludlow, and Babak Heydari are with the School of Systems and Enterprises, Stevens Institute of Technology, Hoboken, NJ, 07030 USA (e-mail: mmosleh@stevens.edu, pludlow@stevens.edu, and babak.heydari@stevens.edu, URL: web.stevens.edu/cens).
}}

\markboth{To APPEAR IN SYSTEMS ENGINEERING  JOURNAL, DOI: \href{http://doi.org/10.1002/sys.21342} {10.1002/sys.21342}}
{Shell \MakeLowercase{\textit{et al.}}: Bare Demo of IEEEtran.cls for Journals}

\maketitle

\begin{abstract}
This paper introduces a framework for studying the interactions of autonomous system components and the design of the connectivity structure in Systems of Systems (SoSs). This framework, which uses complex network models, is also used to study the connectivity structure's impact on resource management. We discuss resource sharing as a mechanism that adds a level of flexibility to distributed systems and describe the connectivity structures that enhance components' access to the resources available within the system. The framework introduced in this paper explicitly incorporates costs of connection and the benefits that are received by direct and indirect access to resources and provides measures of the optimality of connectivity structures. We discuss central and a distributed schemes that, respectively, represent systems in which a central planner determines the connectivity structure and systems in which distributed components are allowed to add and sever connections to improve their own resource access. Furthermore, we identify optimal connectivity structures for systems with various heterogeneity conditions.

\end{abstract}


%
\IEEEpeerreviewmaketitle

\section{Introduction}
%
%
%
%
\IEEEPARstart{S}{ystems} of Systems (SoSs), comprised of heterogeneous components capable of localized, autonomous decision making, are becoming increasingly ubiquitous in a wide range of socio-technical systems \citep{sauser2010systomics,maier1996architecting,jamshidi2011system,mina2006complex}. SoSs often rely on multiple types of localized resources, whose management is a crucial challenge for the optimal performance of the system. SoSs are often operated in highly uncertain environments, because of this it is difficult to anticipate demand for resources in various parts of the system at every moment of time; this means that even if the total demand for a resource can be met, achieving an efficient distribution of the resource is not a trivial challenge. The efficient distribution of resources is, among other factors, a strong function of the system architecture, thus modeling this interdependency---that of the architecture and resource allocation mechanisms---becomes an important area of research in SoSs engineering.

Using a centralized scheme for resource management can be extremely difficult or impossible, because of the large scale, high complexity, and environmental uncertainty of SoSs. 
Attempts to manage all decision making centrally, by gathering information from widely-dispersed system components and then broadcasting decisions back to those components, can lead to a system that is slow to respond to changes in the environment and therefore inefficient (see \cite{koutsopoulos2010auction} for a case in radio systems). Allocation by a central planner is made more complex in situations with heterogeneous system components, as is often the case in SoSs. The tendency of SoSs to have heterogeneous components arises from the fact that these components often operate in different environments, which lead to differing operational constraints and resource requirements. One way to overcome the challenge posed by uncertain variable demand for resources is to ensure that all components of the system are supplied with resources equal to the maximum possible demand for any one component; while this will eliminate the risk of under-supplying any part of the system it is very inefficient and likely to be prohibitively expensive in most systems. Alternatively, a centralized decision making process can allocate resources as and when they are needed throughout the system; however, as was explained before, this can lead to an impractically slow and unresponsive system. 


There has been a shift in the system design paradigm to take advantage of the capabilities that distributed, autonomous or semi-autonomous decision making provides; examples can be seen in many SoSs: fractionated satellite systems\footnote{A fractionated satellite system is a systems architecture concept with the idea being to replace large-scale, expensive, and rigid monolithic satellite systems with a \textit{network} of small-scale, agile, inexpensive, and less complex free-flying satellites that communicate wirelessly and accomplish the same goal as the single monolithic satellite. This new distributed architecture for space systems is argued to be more flexible when responding to uncertainties, such as technology evolution, technical failures, funding availability, and market fluctuations \citep{brown2006value}.} in which detection, processing, and communication tasks are dynamically assigned to members of the satellite cluster \citep{brown2009value, mosleh2014optimal}; communication networks in which frequency spectrum is dynamically allocated for efficient use \citep{mitola1999cognitive, ji2007cognitive}; and groups of unmanned, autonomous vehicles (such as aerial drones) that make dynamic assignment of tasks between them and can each make use of information gathered by other members of the group \citep{alighanbari2005decentralized}. Computational power, bandwidth, and information are examples of scarce resources that the satellites, communication systems, and unmanned vehicles respectively must make efficient use of in their operations.

The distributed, autonomous scheme can also help with optimal resource management of systems of systems: Rather than attempting to address the challenge of resource allocation centrally, one can accept that at any given time some parts of the system will have more resources than needed and other parts fewer; this is not necessarily a problem if the system components are capable of sharing resources between themselves locally. If one part of the system is connected to another part of the system, then those parts are able to exchange resources. These connections could be direct or indirect; for example one part of a system could receive resources from another part via any number of intermediary components. Connections between system components typically come with a price, however, there is most likely some immediate cost associated with creating and maintaining a direct connection between two system components; also, while it may be possible for a resource to be shared indirectly between parts of a system, the quantity or quality of that resource will likely be decreased during the multi-step transmission e.g., attenuation, delay, or cost of involving a third party. 

An architecture perspective, represented by the connectivity structure, can be taken in distributed resource management of a variety of technical and socio-technical SoSs, in which availability of resources is subject to uncertainty. For example, an interconnected network of electrical microgrids can enhance resource access between the units, in which availability of energy resources is affected by the inherent uncertainty of renewable energy resources and fluctuations in electricity demand \citep{katiraei2005micro,saad2011coalitional} i.e., the connectivity structure of the system will play an important role in how unmet demand of one microgrid is supplied by the excess generation of another in an interconnected network of microgrids. Connectivity structure is also a key contributor in distributed resource management of organizations and enterprise systems. For example, in R\&D collaboration networks, firms can either directly combine knowledge, skills, and physical assets to innovate or access innovations of other firms through intermediary firms that serve as conduits through which knowledge and information can spread \citep{konig2012efficiency}. Direct collaboration between two firms has higher benefits, but involves communication and coordination costs while indirect access to resources often discounts benefits due to involving third parties. Given that it is probably inefficient and not practical for every part of a system to be directly connected to every other part, the question becomes that of deciding \textit{what is the best way to connect the system components in order to enhance resource access in uncertain environments}.

Traditional systems engineering methods and theories are not sufficient for analyzing and explaining the dynamics of resource allocation for SoSs with autonomous parts. Any framework that is used to address this challenge has to be able to take into account the local interactions between components of the system while also ensuring that the structure of the connections between components is optimal for the system as a whole. The optimality of the connectivity structure should be evaluated both in the case that it is designed by a central planner as well as when the connectivity structure can change at the discretion of autonomous components.

A viable approach to find the connectivity structures that enhance access to resources within SoSs is to use Network Theory. Network Theory provides methods that go beyond the traditional systems engineering approach as it combines graph theory, game theory, and uncertainty analysis. The system can be modeled as a graph, with the various components of the system being nodes in the graph; the resource-sharing interactions between the autonomous components can be represented using game theory and uncertainty analysis, in the form of games on networks. 

In this paper, we will study the system connectivity structures that enhance access to resources in heterogeneous SoSs. We employ Strategic Network Formation from the economics literature as the underlying framework for finding the optimal connectivity structure when the system is centrally designed, as well as when the connectivity structure is determined dynamically by distributed autonomous components. We discuss the characteristics of those connectivity structures for different heterogeneity conditions.  

The organization of the rest of paper is as follows. In Section~\ref{RS_and_Connectivity}, we discuss a spectrum of systems architectures and explain the role of system connectivity structure and dynamic resource sharing in response to changes in the  environment. In Section~\ref{Modelling_access}, we discuss why Network Theory provides a promising theoretical foundation for studying the architecture of SoSs. In Section~\ref{framework}, we introduce a framework based on Economic Networks to model resource access in SoSs with heterogeneous components. In Section~\ref{OptimalConnectivity}, we introduce models that are used to identify optimal connectivity structures for resource access in SoSs with different heterogeneity conditions, and central- and distributed-design schemes. In Sections~\ref{application},~\ref{discussion}, and~\ref{conclusion}, we discuss applications of the suggested framework, conclude, and provide opportunities for future studies.
%
%
%
%

\section{Resource sharing and system connectivity structure}
\label{RS_and_Connectivity}

Several frameworks have been developed for the architecture of SoSs \citep{maier2009art,rhodes2009architecting,morganwalp2002system}. In this paper, we will focus on using the system's connectivity structure to represent its architecture and will use the framework developed in our previous work \citep{heydari2014,mosleh2015monolithic}. This framework is capable of describing many levels of system connectedness, from fully integral monolithic systems to distributed, adaptive, and dynamic systems. The systems architecture framework is inspired by a general concept of modularity that combines systems modularity \citep{baldwin2000design} and network modularity \citep{newman2006modularity}: that of breaking the larger system into smaller, discrete pieces that are able to interact (communicate) with one another via standardized interfaces \citep{langlois2002modularity}. Given this broad definition of modularity, the systems architecture framework defines five levels of modularity: $M_0-$ Integral (e.g, multi-function valve), $M_1-$ Decomposable (e.g., Smartphone's mainboard), $M_2-$ Modular yet monolithic (e.g., PC's mainboard), $M_3-$ Static-Distributed (e.g., Client-server), and $M_4-$ Dynamic-Distributed (e.g., Internet of Things).

\subsection{Systems architecture spectrum}
The five levels of modularity in the systems architecture framework, developed in our previous work \citep{heydari2014,mosleh2015monolithic}, form a spectrum in which increased modularity improves system responsiveness to the operating environment. The level of modularity, together with systems flexibility, increases from $M_0$ to $M_4$. However, increased modularity comes with increased interfacing costs, increased system complexity, and increased potential for system instability. The operating environment encompasses the physical surroundings of the system and the effects of stakeholder requirements, consumer demand, market forces, policy and regulation, and budgetary constraints. The ability to respond in a flexible manner to all of these environmental factors comes at a cost: if there is little uncertainty in the environment then the flexibility of high modularity will be costly and could lead to instability because of unintended emergent behavior.

The three lowest modularity levels of the framework (i.e., $M_0$, $M_1$, and $M_2$) are related to monolithic systems: systems comprised of a single unit and the interfaces within the monolithic system. The two higher modularity levels of the framework ($M_3$ and $M_4$) correspond to distributed systems that have multiple units capable of inter-unit communications. The interconnected components of the $M_3$ system, which can be clients or servers, communicate and share resources with tasks being assigned to the component with the most appropriate capabilities according to a centralized process.  At the $M_3$ level (``static-distributed") decision-making is centralized and the structure of interactions between components is static; while components in an $M_3$-level system may have different roles, processing capacities, available resources, etc., the assignments do not change over time and the structure of the interactions is fixed. 

While the assignment of tasks to system components is centrally controlled in the static-distributed ($M_3$) architecture, in the dynamic-distributed ($M_4$) architecture tasks are assigned locally to those components that are currently idle or have spare processing capacity for the required task. The assignment decisions are made by the components themselves, i.e., they communicate with each other. The dynamic resource sharing property of an $M_4$ system significantly increases the flexibility and scalability of the system, allowing it to adapt effectively to uncertainties in the environment.

\begin{figure*}[!t]
\centering
\includegraphics[width=5in]{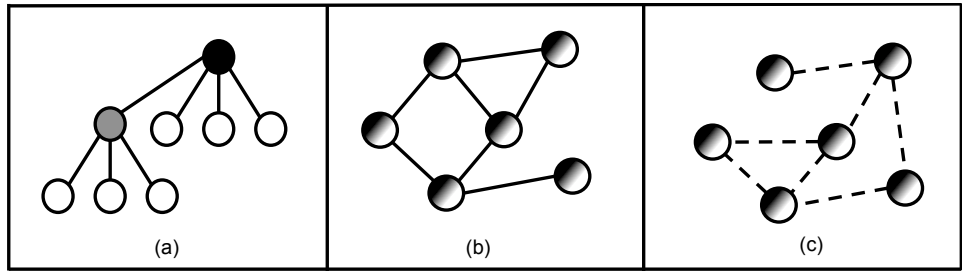}
\caption{Connectivity structure of distributed systems with different levels of flexibility (solid lines represent static connections and dotted lines represent dynamic connections; nodes with solid colors denote fixed roles (client/server) and nodes with gradient color denote components with dynamic roles) (a) Resource allocation and the design of the connectivity structure are centralized ($M_3$). (b) Resource sharing is decentralized but the connectivity structure is static and designed centrally ($M_4$). (c) Resource sharing is decentralized and connectivity structure is dynamic and formed by distributed components ($M_4$). }
\label{dist_arch}
\end{figure*}

While the connectivity structure of a static-distributed ($M_3$) system is typically a tree or two-mode (or bipartite) network, the system connectivity structure of an $M_4$-level dynamic-distributed system will be more complex, having multi-paths and loops. The level of responsiveness to environmental uncertainty of an $M_4$ system can be increased further if its connectivity structure is dynamic, changing in response to environmental factors or additional resource availability. These architectures are illustrated in Figure~\ref{dist_arch}.

\subsection{Multi-layered resource sharing}
\begin{figure}[!t]
\centering
\includegraphics[width=3.5in]{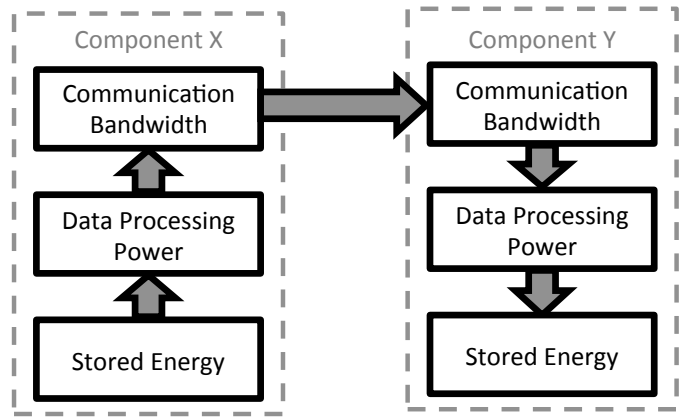}
\caption{An Example of multi-layer resource sharing: a hierarchical multi-layer resource sharing scheme across two satellite systems. While bandwidth sharing is directly possible, sharing of data processing is indirect and is restricted by the limits of bandwidth sharing. Energy sharing is one stage lower and is achieved by moving data processing load to other fractions to save energy. }
\label{MultiLayeredRS}
\end{figure}

The sharing of resources between components of a system that has a dynamic-distributed architecture can be considered to be a multi-layer phenomenon. A multi-layered resource sharing effect occurs when the sharing of a resource by a component affects its consumption of other resources; a component may be able to indirectly access another component's resources through a different resource channel. For example, if one component has excess power supply, it may not be able to directly share power with another component but it could accept a power-consuming task from another component that lacks the power to perform the task. The fractionated satellite system is an example of this scenario as the components of the system have limited local power and processing capacity but the ability to transfer tasks between components via communication channels \citep{brown2009value, mosleh2014optimal}; further examples can be found in cases of distributed computing with heterogeneous hardware and software, and distributed robotic systems \citep{roberts1970computer, wang1994resource}. There are three levels of resource sharing in this case because even though  only data is shared directly, power and processing capacity can also be indirectly shared, as illustrated in Figure~\ref{MultiLayeredRS}. 

The relationship between the layers has this structure because the demand for processing capacity affects the data communications between components, which could have a negative effect on pre-existing tasks requiring communication bandwidth. In addition, a component could delegate a task that has a high associated power drain if its own power supply is at or near capacity.

In a dynamic-distributed system the number of possible configurations for sharing multiple resource types can grow very quickly. The difficulty in optimizing the configuration centrally is one of the primary reasons why in many such dynamic systems the components have some level of autonomy with regard to resource sharing decisions and the connectivity structure itself. Due to the interconnected, dynamic, and autonomous nature of theses systems, the framework required for their analysis has to capture both the component-level autonomous decisions and the effects of the connectivity structure on overall system efficiency.

\section{Modeling resource access in networks}
\label{Modelling_access}
Network theory, an interdisciplinary field at the intersection of computer science, physics, and economics \citep{jackson2008social,easley2010networks,newman2010networks}, provides a promising approach for studying the architecture of SoSs. Network representations make it possible to create a rigorous and domain-independent model of distributed systems. The methods and tools of network theory can be used to study both individual system components' interactions and  aggregate system-level behaviors.

A network, by its very nature, is distributed and can be used to represent system-heterogeneity in the following ways:
\begin{itemize}
\item Degree, centrality, clustering coefficient, and other properties of each node represent a system's structural heterogeneity. 
\item Edge weights in the network represent the heterogeneity in the connections between components in the system. 
\item The type, state, and any associated goals or objective functions of nodes represent heterogeneity of the system components. 
\item Multilayer networks \citep{kivela2014multilayer, de2013centrality} represent the resource heterogeneity (such as energy, information, and risk). 
\item The autonomy of decision-making components in the system can be modeled by considering the network's nodes as agents in a game and using game theory to analyze the autonomous components' behavior. 
\end{itemize}

Although network-based analysis has been used in some systems engineering research, such as when studying product architecture \citep{bartolomei2012engineering, braha2006structure, batallas2006information} and supply chain systems \citep{bellamy2013network}, it has not been used to study resource sharing in systems with a distributed architecture. Different theoretical frameworks, based in network theory, can be used to describe the interactions between autonomous system components depending on the protocol used for making the resource sharing decisions. For example, the interactions can be modeled through exchange networks \citep{bayati2015bargaining, kleinberg2008balanced} if a bargaining process is used to decide on resource sharing actions; in exchange networks the connectivity structure of the network determines each node's \textit{bargaining power} and the way any surplus resources are divided between the nodes. In this paper we will focus on finding the connectivity structure that leads to enhanced resource access by considering two scenarios for the formation of the network connectivity structure: (1) Connectivity structure is static and is determined by a central planner; (2) connectivity structure is dynamic and determined by the autonomous decisions of the distributed system components.

\section{Framework}
\label{framework}
In a system in which components can obtain their required resources both directly and indirectly, deciding which connectivity structure enhances access to resources leads to a dilemma. On one hand, direct connection between two components is costly (e.g., cost of interface); on the other hand, indirect connection may depreciate the benefits of acquiring the resource. Hence, to find the optimal connectivity, we need a framework that explicitly models the heterogeneous costs and benefits of individual components as a function of the network structure (based on the paths of access to the resources). The framework also needs to enable the study of the optimal connectivity structure, and to be able to model and quantify the subsequent trade-offs. 

\subsection{Strategic Networks}
A rigorous framework for studying the optimal connectivity structure is Strategic Network Formation, as it explicitly incorporates the costs and benefits of creating and removing each connection into the model. This framework enables us to study how networks evolve as a result of individual incentives to form links or sever links, and to measure the collective utility of the whole network \citep{jackson2008social}. Hence, this approach is capable of modeling both centralized and autonomous schemes for the formation of the system's connectivity structure. This model was originally introduced in the economics literature and has been widely used to study the economic reasons behind the formation of many real-world networks \citep{jackson2005economics, fricke2012core}.

Most of the theoretical and analytical literature on strategic network formation is built on the work by \cite{jackson1996strategic}. They introduced an economic  network model called the Connection Model in which an agent (node in the network) can benefit from both direct and indirect connections with others, but will only pay a cost for its direct connections. The benefits of indirect connections decrease as the network distance (shortest path) between the nodes increases. This results in a recurring dilemma when creating the optimal connectivity structure (whether static with a central planner or dynamically created by distributed individual agents): (1) should a given agent be connected directly to another agent, in which case they both receive higher benefits, but each also pays a direct connection cost, or (2) should the two nodes be connected through other nodes, in which case, they save the connection cost, but gain only an indirect benefit, which is smaller due to the longer distance between the two nodes. While this dilemma exists for both centrally-designed, static systems and for dynamic systems with autonomous link formation, the resulting structure, in general, can be quite different.

In the Connection Model each agent is assumed to have a utility function, which can represent the costs and benefits of accessing a resource from another part of the system. The notions of \textit{strong efficiency} and \textit{pairwise stability} can represent optimality of the connectivity structures for networks that are built by a central planner and by autonomous components respectively. Strong efficiency means maximizing the total utility of all agents in the network. In other words, for a given set of nodes and utility functions, we say a network is strongly efficient if there is no other network that has higher total utility. Pairwise stability is a generalized form of Nash Equilibrium\footnote{A Nash Equilibrium is a solution concept in game theory for non-cooperative games in which each player is assumed to know the equilibrium strategy of other players and no player can benefit from a unilateral change of strategy if the strategies of others remain unchanged \citep{osborne1994course}.}, which depends on the intention of self-interested individuals to form new links or sever existing ones; a network is said to be pairwise stable if for every pair of nodes: (1) neither has an incentive to sever the link between them if it \textit{does} exist, and (2) only one or zero of them has an incentive to form a link if one \textit{does not} exist \citep{jackson1996strategic}.

\subsection{Connection model}

In this section, we describe the Connection Model as the underlying framework for studying the connectivity structure in order to enhance access to resources in SoS.

For a finite set of agents 
$N=\{1, \dots, n\}$, let $b:\{1, ... ,n-1\} \rightarrow \mathbb{R}$ 
represent the benefit that an agent receives from (direct or indirect) connections to other agents as a function of the distance (shortest path) between them in a graph. Following \cite{jackson1996strategic}, the (distance-based) utility function of each node, $u_i(g)$, in a graph $g$ and the total utility of the graph, $U(g)$, are as follows:

\begin{equation} 
\label{utility_connection_model}
\begin{split}
u_i(g)&=\sum_{j\neq i: j\in N_i^{n-1}(g)} b(d_{ij}(g))-\sum_{j\neq i: j\in N^1_i(g) } c_{ij}\\
U(g)&=\sum_{i=1}^{n}u_i(g)
\end{split}
\end{equation} 

where $N^1_i(g)$ is the set of nodes to which $i$ is linked directly, and $N^k_i(g)$ is the set of nodes that are path-connected to $i$ by a distance no larger than $k$. $d_{ij}(g)$ is the distance (shortest path) between $i$ and $j$, $c_{ij}$ is the cost that node $i$ pays for connecting to $j$, and $b$ is the benefit that node $i$ receives from a connection with another node in the network. We assume that $b(k)>b(k+1)>0$ for any integer $k \geq 1$. 

The $c_{ij}$ values in Equation~\ref{utility_connection_model} are elements of the matrix of potential costs, and only those elements corresponding to direct links will eventually be realized. The connection model has been extended to also account for asymmetry and heterogeneity of benefits (e.g., \cite{persitz2010core}). Note that in the original model introduced by \cite{jackson1996strategic}, it is assumed that the benefits are homogeneous and are a function of the shortest path between two nodes, while direct connection costs can be heterogeneous in general.   We will revisit this later in the paper in Section~\ref{Dynamic_heterogeneous_connectivity}. However, even only assuming cost heterogeneity can capture many real forms of complexities that arise, from having agents with different bandwidths and information processing capacities, to distance-based cost variations. Moreover, heterogeneous cost models automatically capture heterogeneity in direct benefits, since the difference in direct benefits can be absorbed into the cost.

Potential costs and benefits are identified based on components' characteristics, such as location \citep{johnson2003spatial}, available energy or processing power, and interface standards. The assumption is that the states and attributes of the nodes are known and are inputs to the model. Hence, this model does not optimize the location of nodes, or other attributes that are related to individual nodes. Instead, it is used to study which components should be connected to each other in order to fulfill a system-level criterion.

For systems with a centrally-determined connectivity structure, we use the notion of an efficient network, that is the network structure that maximizes total utility of all nodes: 

Let the complete graph $g^N$ denote the set of all subsets of $N$ of size 2. The network $\tilde{g}$ is efficient, if $U(\tilde{g}) \geq U(g^\prime)$ for all $g^\prime \subset g^N$, which indicates that:
\begin{equation} 
\tilde{g}=\argmax_g\sum_{i=1}^{n}u_i(g)
\end{equation}

For systems where autonomous components are allowed to change the structure, using strong efficiency as the sole notion of optimality is not sufficient. In such systems, different components can change the structure based on local incentives, which might or might not be aligned with the global optimal efficiency. The concept of optimality for the connectivity structure of systems with autonomous components can be defined based on a game-theoretic equilibrium that captures individual and mutual incentives for the formation of connections. Hence, we will use the notion of pairwise stability as defined by \cite{jackson1996strategic} and that has been used in many other subsequent works. This definition describes the intuitive scenario in which adding a link between two agents requires a mutual decision while decisions to remove links can be unilateral.
The network $g$ is \textit{pairwise stable} if:

\begin{enumerate}[(i)]
\item for all $ij \in g$, $u_i(g) \geq u_i(g - ij)$ and $u_j(g) \geq u_j(g - ij)$ and
\item for all $ij \in g$, if $u_i(g + ij) \geq u_i(g)$ then $u_j(g + ij) < u_j(g)$.
\end{enumerate}

where $g+ij$ denotes the network obtained by adding link $ij$ to the existing network $g$, and $g-ij$ represents the network obtained by removing link $ij$ from the existing network $g$.

The connection model captures dependencies of components and synergies at the micro level. The utility function of each component represents its goals, which can be aligned or not aligned with those of the whole system. The utility function depends on the connections of one component to the others and can account for the heterogeneous states of components. The utility function has a general form and can capture non-linearity in the preference functions of autonomous components. 

Thus far, we have mainly discussed (strong) efficiency and pairwise stability as two system-level criteria. However, depending on the context, a variety of criteria can be defined to measure the performance of the system based on individual components' utility functions. The notion of (strong) efficiency is defined based on the assumption that a central authority would design a system to maximize the sum of individual utilities. One can also consider Pareto efficiency as a criterion for a centrally-designed system. However, the pairwise stability metric can represent ``overall satisfaction" in the sense that no autonomous component in the system would be willing to change its connections, as this would not improve its utility. The assumption behind the pairwise stability or two-sided link formation is that a link is formed upon the ``mutual consent" of two agents. However, one can study the connectivity structure that results from one-sided and non-cooperative link formation, where agents unilaterally decide to form the links with another agent \citep{bala2000noncooperative}.\footnote{For a thorough comparison between strong efficiency, Pareto efficiency, and pairwise stability, please refer to \cite{jackson2008social}, Chapter 6, Section 2.}

Using the connection model framework we can find the optimal connectivity structure for various conditions for costs and benefits associated with access to a resource in the system.

\begin{figure}[!t]
\centering
\includegraphics[width=3.3in]{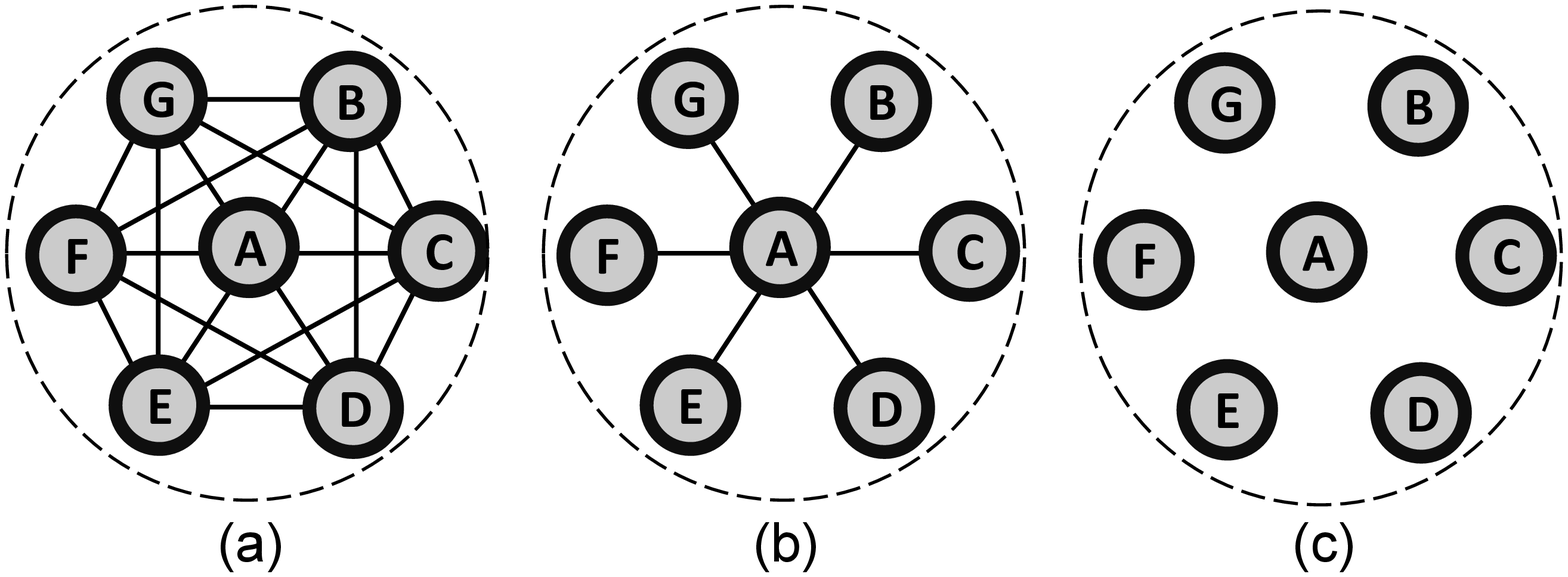}
\caption{Optimal connectivity structure for optimized resource access when the cost of connection between components is homogeneous and the connectivity structure is designed centrally. (a) Low cost of connection i.e., $c < b(1) -b(2)$. (b) Moderate cost of connection i.e., $b(1) - b(2) < c < b(1) + 0.5(n - 2)b(2)$. (c) High cost of connection i.e., $c > b(1) + 0.5(n - 2)b(2)$.}
\label{HomoNetwork}
\end{figure}

\section{Optimal connectivity structure for resource access}
\label{OptimalConnectivity}
\subsection{Homogeneous connection cost}

A system in which connecting every two components has equal cost can be presented by the simple homogeneous form of Equation~\ref{utility_connection_model}, where $c_{ij}=c$. Following~\cite{bloch2007formation}, when the connectivity structure is decided by a central planner, the optimal network does not have a diameter greater than two and will have the following structures depending on the cost and the benefit function\footnote{For details of mathematical proofs, please refer to \cite{jackson2008social} Chapter 6, Section 3.}: 
\begin{enumerate}[(i)]
\item a complete graph if $b(1)-b(2)>c$,
\item a star structure if $b(1)-b(2)<c<b(1)+0.5(n-2)b(2)$,
\item an empty graph if $c>b(1)+0.5(n-2)b(2)$.
\end{enumerate}
The structures of efficient networks imply that when the cost of connecting two components in the system is below a certain limit, it is worthwhile to connect all components so that they benefit from direct access to each others resources. However, for a moderate cost of connection, a star structure optimizes access; in this structure a component acts as a hub through which other components can access resources from throughout the system via at most one intermediary. For this cost range, star is the unique efficient structure in that it has the minimum number of links connecting all nodes and minimizes the average path length given the minimal number of links. When the connection cost is beyond a certain limit, sharing resources is not beneficial in the system. These structures are depicted in Figure~\ref{HomoNetwork}.

In a system in which components can autonomously establish and sever links to maximize their own access to resources, the optimal network is not necessarily unique. Following \cite{bloch2007formation} the description of the pairwise stable networks with homogeneous costs is as follows:

\begin{enumerate}[(i)]
\item for $c<b(1)-b(2)$, the unique pairwise stable network is the complete graph,
\item for $b(1)-b(2)<c < b(1) $, a star structure is pairwise stable, but not necessarily the unique pairwise stable graph ,
\item for $b(1)<c$, any pairwise stable network which is non-empty is such that each player has at least two links and thus is inefficient.
\end{enumerate}

Although for the low connection cost the efficient and pairwise networks coincide, for the higher costs, the stable network is not unique and may not be the same as the efficient network. It is desirable to know how much total inefficiency will result from allowing networks to form at the discretion of autonomous components as opposed to being designed by a central planner. Knowing the magnitude of this inefficiency is often recognized as the \textit{price of anarchy} in the literature as was first introduced and coined by \cite{papadimitriou2001algorithms}. 
%

\subsection{Heterogeneous connection cost}

The homogeneity assumption does not hold in many real-world systems, where the cost of connection is different from one link to another. A number of models have been proposed in the literature to introduce heterogeneity into the connection model \citep{galeotti2006network, jackson2005economics, vandenbossche2013network}. As an example of these heterogeneous models, we focus on the Separable Connection Cost model \citep{heydari2015efficient}, which is motivated by the distributed systems in which heterogeneous components are each endowed with some \textit{budget} and the total budget needed to establish and maintain connections for each component can be approximated to be proportional to the number of components to which it is connected. In this model each node pays a fixed cost for each connection independent of to whom it connects (i.e., $c_{ij}=c_i$ in Equation~\ref{utility_connection_model}), but this cost varies from node to node.

When centrally designed, the connectivity structure that optimizes access to resources with separable and heterogeneous connection costs is as follows (mathematical proofs are provided in \cite{heydari2015efficient} )
\begin{itemize}
\item Assuming that $c_1<c_2<\dots<c_n$, let $m$ be the largest integer between 1 and $n$ such that $2b(1) + 2(m-2)b(2) > (c_m + c_1)$. 
\item If $i > m$, then $i$ is isolated. If $i \leq m$, then there is exactly one link between $i$ and $1$; 
\item also there is one link between $i$ and $j (1 < i,j \leq m)$ if and only if $b(1)-b(2) > 0.5(c_i + c_j)$.

\end{itemize}

\begin{figure}[!t]
\centering
\includegraphics[width=1.7in]{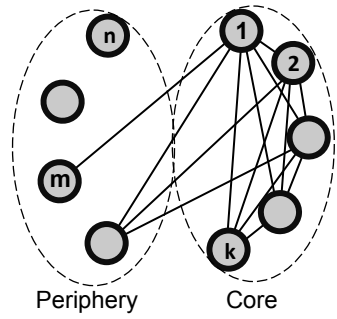}
\caption{Optimal connectivity structure for optimized resource access when the cost of connection between components is heterogeneous and separable, and connectivity structure is designed centrally.}
\label{GeneralizedStar}
\end{figure}

In the efficient connectivity structure, components with high connection cost are isolated and the rest of the components are connected in a \textit{generalized star} structure. In this structure the component with the minimum connection cost plays the role of the hub, through which other components can access each other's resources. Moreover, if the cost of connection between two components is less than the gain in benefit of a direct connection compared to indirect, they are also connected. This will form a \textit{Core-Periphery} structure where components in the Core are fully interconnected and the components in the Periphery are only connected to those in the Core (Figure~\ref{GeneralizedStar}). Although benefits are still assumed to be homogeneous, one can easily take into account heterogeneity of direct benefits through cost, as long as the separability assumption is maintained, i.e. cost and direct benefit terms appear together in all analyses and the cost terms can capture heterogeneity of direct benefits by embedding them as an offset to the fixed costs of nodes.

\subsection{Dynamic heterogeneous connectivity}
\label{Dynamic_heterogeneous_connectivity}

By integrating heterogeneity of the environment and components' characteristics (e.g., processing capacity, state) into the model, we can capture their effects on the dynamic interactions of the autonomous system components that evolve the connectivity structure. Based on the connection model and agent-based simulation, \cite{heydari2015emergence} suggest a computational framework for studying the connectivity structure that emerges from the component-level decisions for creating and severing links. This model extends the original model of \cite{jackson1996strategic} to capture the effect of both heterogeneous benefits and heterogeneous connection costs on the pairwise stable network. Note that in this model, due to the heterogeneity in both benefits and costs of connections, finding the efficient network is intractable in general. Using this model, self-optimizing components can play a network formation game in a heterogeneous environment and organize themselves in a manner that balances the benefits of access to resources against the associated costs in a way that also takes into account the limited processing capacity of the components.

Based on the cost and benefits of access to a resource defined in this model, each component maximizes its own utility by establishing new links, with the mutual consent of the components at the other end of those links, or unilaterally removing existing ones. In a heterogeneous environment, an autonomous component is faced with a fundamental dilemma regarding the aggregate heterogeneity of its connections. On one hand, maximizing the diversity of connections, i.e., direct and indirect, is desirable because it ensures access to a larger pool of resources to respond to changes of environment. On the other hand, each component, when considered to have limited processing capacity, can only handle a certain level of heterogeneity in its direct connections. The reason is that each link imposes a transaction cost on the connected nodes that is a function of expected heterogeneity of the link's endpoints. The effect of the environment further amplifies this dilemma. This is because more heterogeneous environments give rise to a higher expected benefit to nodes from a given diversity in their connections. 

In this model, heterogeneity of the system environment is captured by the nodes' states. That is, each node in the network exchanges resources with a different environment, which influences its state. Another aspect of this model is the link formation capacity, which is a characteristic of each node. Each connection imposes a cost on a node and the node cannot maintain connections that have a total cost more than its capacity. The cost of link formation depends on the internal states of the two connected nodes. This implies that it is more expensive for an autonomous component to connect to another component that is very different compared with connecting to a component with similar characteristics. For instance, in communication networks, direct connection to a distant node is more expensive than connecting to a node in a close neighborhood. A node's state also affects the benefits another node receives from connecting to it. Having a path to a node with different characteristics provides greater opportunities for resource exchange. For instance, in the communication network a connection to a distant node provides access to a new geographic location.

\begin{figure}[!t]
\centering
\includegraphics[width=3.5in]{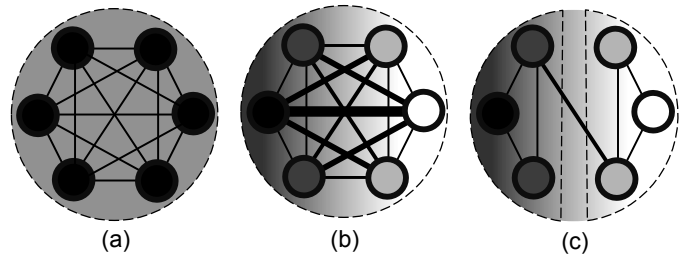}
\caption{Optimal connectivity structure that has emerged from formation and removal of links by autonomous components, which seek to improve their access to resources in the system \citep{heydari2015emergence}. The thickness of the links denotes the connection cost, which is a function of the difference between nodes' states (a) Homogeneous set of components in a homogeneous environment. (b) In a heterogeneous environment, nodes have different rates of resource exchange with the environment and will become heterogeneous over time. The transaction cost of having many links will increase as a result. (c) Due to the limited processing capacities, components cannot afford all of their links, and sever a large percentage of them to keep their total transaction costs below their capacity, while still having access to a diverse set of nodes. This creates modularity in the connectivity structure.}
\label{tem}
\end{figure}

\begin{figure}[!t]
\centering
\includegraphics[width=3.5in]{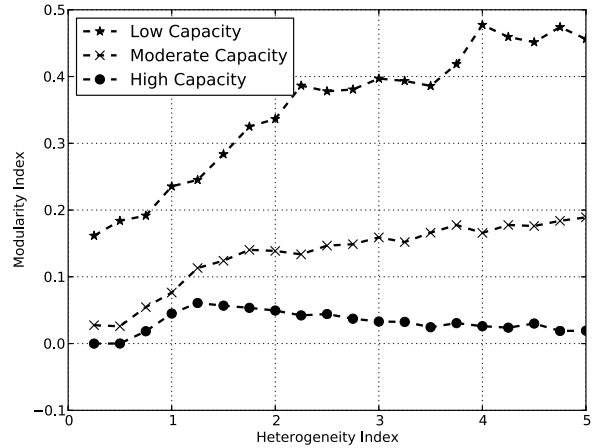}
\caption{Effect of environmental heterogeneity on modularity index of optimal connectivity structure when autonomous components create and sever links to improve their access to heterogeneous resources within the system \citep{heydari2015emergence}.}
\label{TEM_het_mod}
\end{figure}

The pairwise stable network that is formed based on decisions of individual heterogeneous components is not unique. This makes the analysis of the exact connectivity structure challenging, particularly when the network is large. However, the study of structural features reveals that the pairwise stable connectivity structures exhibit distinctive characteristics for systems containing self-optimizing heterogeneous components.
Intuitively, \textit{modular communities} \citep{newman2006modularity} emerge when autonomous components maximize their indirect connections' diversity while keeping their link cost within their processing capacities. This is achieved by obtaining indirect benefits through direct connections to components with higher processing capacities that have the ability to manage a larger number of direct connections to heterogeneous resources. Figure~\ref{tem} illustrates how connectivity structures evolve as the result of self-optimizing decisions in creating and severing connections. To measure the strength of the community structure, \cite{heydari2015emergence} used the modularity index $Q$ developed by \cite{newman2004finding}, where $Q=1$ is the maximum and indicates strongest community structure. The results in \citep{heydari2015emergence} show that when heterogeneity of the environment (measured by diversity of nodes' states) is low and components have high processing capacities, the connectivity structure has a lower modularity index. However, high environmental heterogeneity together with limited processing capacities results in a higher modularity index (Figure~\ref{TEM_het_mod}).

Note that although the changes of connectivity structure by autonomous agents in real-time might be partially attributed to the operation of the system, the proposed dynamic network formation model can be used as the basis of several architectural decisions. The model can be used to determine the initial topology of an autonomous system based on a given environment profile. The proposed framework can also be employed to decide about the level of autonomy of distributed agents, i.e., which agents are allowed to dynamically form or sever links (and with whom). Moreover, using the framework, one can decide the initial distribution of resources and the allocation of heterogeneous agents in the network to influence the agents' decisions on link formation.

\section{A note on potential applications}
\label{application}
The proposed framework is applicable in determining the connectivity structure of SoSs when components can autonomously share resources in order to manage uncertainty in the availability of distributed resources. This includes technical and socio-technical systems such as the Internet of Things (IoT), Connected Autonomous Vehicles, fractionated satellite systems, R\&D collaboration networks, or hybrid teams of human and autonomous agents for disaster response. The main focus of this paper is on introducing a framework to enhance resource access in SoS and expanding on the theoretical foundations. In this section we will discuss two potential application areas for the framework. The finer details of these implementations are beyond the scope of the present paper and require that one quantifies the connection costs and benefit functions in the context of the problem, captures components' heterogeneous characteristics in the individual agents' utility functions, and uses appropriate system-level criteria to determine the connectivity structure.

In fractionated satellite systems, multi-layer resource sharing enables the exchanging of resources, such as computational capacity, energy, and communication bandwidth, across fractions in the face of uncertainty in the availability of resources. The sources of uncertainty include variations in demand (e.g. market fluctuations and changes of stakeholders' requirements) and supply (e.g., change of mission and technical failure). It is neither practical nor efficient for all fractions to communicate directly with each other, thus it becomes important to find the communications connectivity structure between fractions that optimally enhances resource access throughout the system. This can be modeled through the proposed framework where nodes represent satellite fractions and nodes' states capture the fraction's heterogeneous characteristics (e.g., processing capacity limit and locations). Connection costs and the benefits of direct/indirect resource access can be defined as a function of nodes' states. 

The proposed model can also be employed to study the effect of connectivity structure on performance of socio-technical systems such as hybrid teams of human and autonomous agents. Many critical systems of the future will rely on hybrid teams, in which human and autonomous technology agents (such as autonomous robots, self-driving cars or autonomous micro-grids) coordinate their actions, cooperate, share information, and dynamically divide sensing, information processing, and decision-making tasks. For example in a disaster response scenario, a group of geographically distributed heterogeneous agents need to cooperate and share information in a rapidly changing and uncertain environment. On the one hand agents seek to improve their access to information while their processing capacity is limited in handling connections. On the other hand, receiving information through intermediaries is subject to delay and noise. An extended model based on the dynamic network formation model (Section~\ref{Dynamic_heterogeneous_connectivity}) can be used to study connectivity structures that result in a stable network in which agents---while having autonomy over connection formation or severance---do not see it beneficial to deviate from the designated structure. 

\section{Discussion}
\label{discussion}
The framework proposed in this paper is domain independent and can be applied in a variety of contexts to study the connectivity structure of systems of systems comprised of heterogeneous and autonomous components. The framework offers a new perspective on distributed resource management in SoS under uncertainty that has been missing in the existing literature. However, the proposed framework is not intended to replace existing approaches that focus on reliability, or context-dependent operational or functional models. Instead, the proposed model can complement the existing approaches for resource management in SoS. Integrating an architecture perspective approach into existing frameworks is a topic of future research.

The key difference between the proposed economic network model and classical operations research network approaches, such as minimum spanning trees \citep{kruskal1956shortest}, is the ability to capture autonomous behavior of heterogeneous components. In the proposed framework the utility function of individual components has a general form and, together with the concepts of efficiency and pairwise stability, can be used to study both central and decentralized schemes for forming a connectivity structure. Moreover, the framework explicitly incorporates the benefits of connections as a function of the distance between  components and accounts for heterogeneous connection costs. The suggested framework can be used to study how a connectivity structure emerges within a group of agents that are improving their own utilities by severing and creating links (e.g., a communication network of autonomous agents for disaster response). The model enables us to study the economic reasons behind emergence of network structures as a result of individual components' decisions, and also provides us with insights to steer the evolution of those structures by influencing individuals' incentives. In contrast, a minimum spanning tree approach might be used to centrally design a cost effective network encompassing all components in a system (e.g., laying out cables for a telecommunication networks in a new area \citep{graham1985history}). 

The proposed framework can be used to find the optimal network topology for a given set of parameters at a moment in time. Once a new component  is added to/removed from the system, the same framework can be used to find the new optimal topology. However, the proposed framework does not capture the optimal transition strategy and the required changes in the overall architecture to obtain a globally optimal network. This depends on a set of parameters, such as the expected frequency of addition/removal, and the location and interdependency of added/removed nodes that are not considered in this paper. Integrating optimal strategies for transitions in systems with dynamic set of components and finding a global optimal topology are important directions that can complement this work.

We used deterministic cost and benefit functions in the optimal connectivity structure models in this paper. When using stochastic functions, with expected values of costs and benefits, similar results will still be valid. Using stochastic functions for costs and benefits enables the integration of other component characteristics such as reliability into the model, i.e., the probability of failure of each component will negatively affect the expected benefits that are received from connections to that node. However, for more complex analysis, one needs to modify the framework to accommodate probability distributions of cost and benefit functions.


This paper focused mainly on enhancing individual components' access to resources within the system by finding an optimal connectivity structure. However, the study of mechanisms for sharing resources between autonomous components (a.k.a. Multi-Agent Resource Allocation) is another topic, which is widely studied jointly by computer scientists and economists. These mechanisms are intended to align individual components' utilities, obtained from sharing a resource, with system-wide goals. The resource sharing mechanisms between system components can be defined according to a variety of protocols depending on factors such as the type of the resource (e.g., single vs. multi-unit, continuous vs. discrete), and complexity of the resource allocation algorithm. Many of these protocols are inspired by market mechanisms such as auctions and negotiation \citep{chevaleyre2006issues}.

\section{Conclusion}
\label{conclusion}
Dynamic resource sharing, as a systems mechanism,  can add a level of flexibility to SoSs and improve their responsiveness to uncertainty in the environment. In this paper, we took a systems architecture approach to distributed resource management in SoSs. We introduced a framework based on Economic Networks for the connectivity structure of SoSs in which components can share resources through direct and indirect connections. This framework enables us to study the effect of the connectivity structure on individual components' utility that is obtained from access to diverse resources available to other components.



The optimal connectivity structure depends on the heterogeneity parameters of the system, the environment, and the way in which the connectivity structure is formed (i.e., by a central planner or distributed components). The proposed model explicitly incorporates the cost of creating and maintaining a connection between two components as well as the benefits that are received through direct and indirect access to a resource. It can also capture a wide range of heterogeneity of system parameters and the environment. Moreover, the notion of strong efficiency is used to represent the optimality of a connectivity structure created by a central planner; similarly, the notion of pairwise stability is used to study the structures emerging from self-optimizing components' incentives to create and sever links. 



In this paper, we mainly focused on the optimal connectivity structure of few particular heterogeneity conditions. However, the cost and benefit functions in the proposed framework can be extended to capture various levels of heterogeneity in distributed systems while finding the optimal network remains fairly tractable. For example, systems where constituents can be divided into a number of groups (islands), in which connections between islands are generally more costly than connections within islands, can be studied based on the Island-connection model \citep{jackson2005economics}. Moreover, in the original model the benefit received from connection to another component is a function of the distance between two components. However, the benefits of resource access might be negatively affected by the number of connections to the component providing the resource. Extended models such as the degree-distance-based connections model \citep{mohlmeier2013degree} can be used to model this effect.

\section*{Acknowledgment}
This work was supported in part by DARPA/NASA Ames Contract Number: NNA11AB35C and INCOSE/SERC developing Theory of Systems Engineering.

\ifCLASSOPTIONcaptionsoff
  \newpage
\fi



\bibliographystyle{IEEEtranN}
\bibliography{ResourceSharing}

\begin{thebibliography}{53}
\providecommand{\natexlab}[1]{#1}
\providecommand{\url}[1]{#1}
\csname url@samestyle\endcsname
\providecommand{\newblock}{\relax}
\providecommand{\bibinfo}[2]{#2}
\providecommand{\BIBentrySTDinterwordspacing}{\spaceskip=0pt\relax}
\providecommand{\BIBentryALTinterwordstretchfactor}{4}
\providecommand{\BIBentryALTinterwordspacing}{\spaceskip=\fontdimen2\font plus
\BIBentryALTinterwordstretchfactor\fontdimen3\font minus
  \fontdimen4\font\relax}
\providecommand{\BIBforeignlanguage}[2]{{%
\expandafter\ifx\csname l@#1\endcsname\relax
\typeout{** WARNING: IEEEtranN.bst: No hyphenation pattern has been}%
\typeout{** loaded for the language `#1'. Using the pattern for}%
\typeout{** the default language instead.}%
\else
\language=\csname l@#1\endcsname
\fi
#2}}
\providecommand{\BIBdecl}{\relax}
\BIBdecl

\bibitem[Sauser et~al.(2010)Sauser, Boardman, and Verma]{sauser2010systomics}
B.~Sauser, J.~Boardman, and D.~Verma, ``Systomics: Toward a biology of system
  of systems,'' \emph{Systems, Man and Cybernetics, Part A: Systems and Humans,
  IEEE Transactions on}, vol.~40, no.~4, pp. 803--814, 2010.

\bibitem[Maier(1996)]{maier1996architecting}
M.~W. Maier, ``Architecting principles for systems-of-systems,'' in
  \emph{INCOSE International Symposium}, vol.~6, no.~1.\hskip 1em plus 0.5em
  minus 0.4em\relax Wiley Online Library, 1996, pp. 565--573.

\bibitem[Jamshidi(2011)]{jamshidi2011system}
M.~Jamshidi, \emph{System of systems engineering: innovations for the
  twenty-first century}.\hskip 1em plus 0.5em minus 0.4em\relax John Wiley \&
  Sons, 2011, vol.~58.

\bibitem[Mina et~al.(2006)Mina, Braha, and Bar-Yam]{mina2006complex}
A.~A. Mina, D.~Braha, and Y.~Bar-Yam, ``Complex engineered systems: A new
  paradigm,'' in \emph{Complex engineered systems}.\hskip 1em plus 0.5em minus
  0.4em\relax Springer, 2006, pp. 1--21.

\bibitem[Koutsopoulos and Iosifidis(2010)]{koutsopoulos2010auction}
I.~Koutsopoulos and G.~Iosifidis, ``Auction mechanisms for network resource
  allocation,'' in \emph{Modeling and Optimization in Mobile, Ad Hoc and
  Wireless Networks (WiOpt), 2010 Proceedings of the 8th International
  Symposium on}.\hskip 1em plus 0.5em minus 0.4em\relax IEEE, 2010, pp.
  554--563.

\bibitem[Brown and Eremenko(2006)]{brown2006value}
O.~Brown and P.~Eremenko, ``The value proposition for fractionated space
  architectures,'' DTIC Document, Tech. Rep., 2006.

\bibitem[Brown et~al.(2009)Brown, Eremenko, and Collopy]{brown2009value}
O.~C. Brown, P.~Eremenko, and P.~D. Collopy, \emph{Value-centric design
  methodologies for fractionated spacecraft: progress summary from phase 1 of
  the DARPA system F6 program}.\hskip 1em plus 0.5em minus 0.4em\relax Defense
  Technical Information Center, 2009.

\bibitem[Mosleh et~al.(2014)Mosleh, Dalili, and Heydari]{mosleh2014optimal}
M.~Mosleh, K.~Dalili, and B.~Heydari, ``Optimal modularity for fractionated
  spacecraft: The case of system f6,'' \emph{Procedia Computer Science},
  vol.~28, pp. 164--170, 2014.

\bibitem[Mitola~III and Maguire~Jr(1999)]{mitola1999cognitive}
J.~Mitola~III and G.~Q. Maguire~Jr, ``Cognitive radio: making software radios
  more personal,'' \emph{Personal Communications, IEEE}, vol.~6, no.~4, pp.
  13--18, 1999.

\bibitem[Ji and Liu(2007)]{ji2007cognitive}
Z.~Ji and K.~Liu, ``Cognitive radios for dynamic spectrum access-dynamic
  spectrum sharing: A game theoretical overview,'' \emph{Communications
  Magazine, IEEE}, vol.~45, no.~5, pp. 88--94, 2007.

\bibitem[Alighanbari and How(2005)]{alighanbari2005decentralized}
M.~Alighanbari and J.~P. How, ``Decentralized task assignment for unmanned
  aerial vehicles,'' in \emph{Decision and Control, 2005 and 2005 European
  Control Conference. CDC-ECC'05. 44th IEEE Conference on}.\hskip 1em plus
  0.5em minus 0.4em\relax IEEE, 2005, pp. 5668--5673.

\bibitem[Katiraei et~al.(2005)Katiraei, Iravani, and Lehn]{katiraei2005micro}
F.~Katiraei, M.~Iravani, and P.~W. Lehn, ``Micro-grid autonomous operation
  during and subsequent to islanding process,'' \emph{Power Delivery, IEEE
  Transactions on}, vol.~20, no.~1, pp. 248--257, 2005.

\bibitem[Saad et~al.(2011)Saad, Han, and Poor]{saad2011coalitional}
W.~Saad, Z.~Han, and H.~V. Poor, ``Coalitional game theory for cooperative
  micro-grid distribution networks,'' in \emph{Communications Workshops (ICC),
  2011 IEEE International Conference on}.\hskip 1em plus 0.5em minus
  0.4em\relax IEEE, 2011, pp. 1--5.

\bibitem[K{\"o}nig et~al.(2012)K{\"o}nig, Battiston, Napoletano, and
  Schweitzer]{konig2012efficiency}
M.~D. K{\"o}nig, S.~Battiston, M.~Napoletano, and F.~Schweitzer, ``The
  efficiency and stability of r\&d networks,'' \emph{Games and Economic
  Behavior}, vol.~75, no.~2, pp. 694--713, 2012.

\bibitem[Maier(2009)]{maier2009art}
M.~W. Maier, \emph{The art of systems architecting}.\hskip 1em plus 0.5em minus
  0.4em\relax CRC press, 2009.

\bibitem[Rhodes et~al.(2009)Rhodes, Ross, and
  Nightingale]{rhodes2009architecting}
D.~H. Rhodes, A.~M. Ross, and D.~J. Nightingale, ``Architecting the system of
  systems enterprise: Enabling constructs and methods from the field of
  engineering systems,'' in \emph{Systems Conference, 2009 3rd Annual
  IEEE}.\hskip 1em plus 0.5em minus 0.4em\relax IEEE, 2009, pp. 190--195.

\bibitem[Morganwalp and Sage(2002)]{morganwalp2002system}
J.~Morganwalp and A.~P. Sage, ``A system of systems focused enterprise
  architecture framework and an associated architecture development process,''
  \emph{Information-Knowledge-Systems Management}, vol.~3, no. 2-4, pp.
  87--105, 2002.

\bibitem[Heydari et~al.(2016)Heydari, Mosleh, and Dalili]{heydari2014}
B.~Heydari, M.~Mosleh, and K.~Dalili, ``From modular to distributed open
  architectures: A unified decision framework,'' \emph{Systems Engineering},
  vol. In Press, 2016.

\bibitem[Mosleh et~al.(2016)Mosleh, Dalili, and Heydari]{mosleh2015monolithic}
M.~Mosleh, K.~Dalili, and B.~Heydari, ``Distributed or monolithic? a
  computational architecture decision framework,'' \emph{IEEE Systems journal},
  vol. In Review, 2016.

\bibitem[Baldwin and Clark(2000)]{baldwin2000design}
C.~Y. Baldwin and K.~B. Clark, \emph{Design rules: The power of
  modularity}.\hskip 1em plus 0.5em minus 0.4em\relax MIT press, 2000, vol.~1.

\bibitem[Newman(2006)]{newman2006modularity}
M.~E. Newman, ``Modularity and community structure in networks,''
  \emph{Proceedings of the National Academy of Sciences}, vol. 103, no.~23, pp.
  8577--8582, 2006.

\bibitem[Langlois(2002)]{langlois2002modularity}
R.~N. Langlois, ``Modularity in technology and organization,'' \emph{Journal of
  economic behavior \& organization}, vol.~49, no.~1, pp. 19--37, 2002.

\bibitem[Roberts and Wessler(1970)]{roberts1970computer}
L.~G. Roberts and B.~D. Wessler, ``Computer network development to achieve
  resource sharing,'' in \emph{Proceedings of the May 5-7, 1970, spring joint
  computer conference}.\hskip 1em plus 0.5em minus 0.4em\relax ACM, 1970, pp.
  543--549.

\bibitem[Wang and Premvuti(1994)]{wang1994resource}
J.~Wang and S.~Premvuti, ``Resource sharing in distributed robotic systems
  based on a wireless medium access protocol (csma/cd-w),'' in
  \emph{Intelligent Robots and Systems' 94.'Advanced Robotic Systems and the
  Real World', IROS'94. Proceedings of the IEEE/RSJ/GI International Conference
  on}, vol.~2.\hskip 1em plus 0.5em minus 0.4em\relax IEEE, 1994, pp. 784--791.

\bibitem[Jackson et~al.(2008)]{jackson2008social}
M.~O. Jackson \emph{et~al.}, \emph{Social and economic networks}.\hskip 1em
  plus 0.5em minus 0.4em\relax Princeton University Press Princeton, 2008,
  vol.~3.

\bibitem[Easley and Kleinberg(2010)]{easley2010networks}
D.~Easley and J.~Kleinberg, \emph{Networks, crowds, and markets: Reasoning
  about a highly connected world}.\hskip 1em plus 0.5em minus 0.4em\relax
  Cambridge University Press, 2010.

\bibitem[Newman(2010)]{newman2010networks}
M.~Newman, \emph{Networks: an introduction}.\hskip 1em plus 0.5em minus
  0.4em\relax Oxford University Press, 2010.

\bibitem[Kivel{\"a} et~al.(2014)Kivel{\"a}, Arenas, Barthelemy, Gleeson,
  Moreno, and Porter]{kivela2014multilayer}
M.~Kivel{\"a}, A.~Arenas, M.~Barthelemy, J.~P. Gleeson, Y.~Moreno, and M.~A.
  Porter, ``Multilayer networks,'' \emph{Journal of Complex Networks}, vol.~2,
  no.~3, pp. 203--271, 2014.

\bibitem[De~Domenico et~al.(2013)De~Domenico, Sol{\'e}-Ribalta, Omodei,
  G{\'o}mez, and Arenas]{de2013centrality}
M.~De~Domenico, A.~Sol{\'e}-Ribalta, E.~Omodei, S.~G{\'o}mez, and A.~Arenas,
  ``Centrality in interconnected multilayer networks,'' \emph{arXiv preprint
  arXiv:1311.2906}, 2013.

\bibitem[Bartolomei et~al.(2012)Bartolomei, Hastings, de~Neufville, and
  Rhodes]{bartolomei2012engineering}
J.~E. Bartolomei, D.~E. Hastings, R.~de~Neufville, and D.~H. Rhodes,
  ``Engineering systems multiple-domain matrix: An organizing framework for
  modeling large-scale complex systems,'' \emph{Systems Engineering}, vol.~15,
  no.~1, pp. 41--61, 2012.

\bibitem[Braha and Bar-Yam(2006)]{braha2006structure}
D.~Braha and Y.~Bar-Yam, ``The structure and dynamics of complex product
  design,'' in \emph{Complex Engineered Systems}.\hskip 1em plus 0.5em minus
  0.4em\relax Springer, 2006, pp. 40--71.

\bibitem[Batallas et~al.(2006)Batallas, Yassine,
  et~al.]{batallas2006information}
D.~A. Batallas, A.~Yassine \emph{et~al.}, ``Information leaders in product
  development organizational networks: Social network analysis of the design
  structure matrix,'' \emph{Engineering Management, IEEE Transactions on},
  vol.~53, no.~4, pp. 570--582, 2006.

\bibitem[Bellamy and Basole(2013)]{bellamy2013network}
M.~A. Bellamy and R.~C. Basole, ``Network analysis of supply chain systems: A
  systematic review and future research,'' \emph{Systems Engineering}, vol.~16,
  no.~2, pp. 235--249, 2013.

\bibitem[Bayati et~al.(2015)Bayati, Borgs, Chayes, Kanoria, and
  Montanari]{bayati2015bargaining}
M.~Bayati, C.~Borgs, J.~Chayes, Y.~Kanoria, and A.~Montanari, ``Bargaining
  dynamics in exchange networks,'' \emph{Journal of Economic Theory}, vol. 156,
  pp. 417--454, 2015.

\bibitem[Kleinberg and Tardos(2008)]{kleinberg2008balanced}
J.~Kleinberg and {\'E}.~Tardos, ``Balanced outcomes in social exchange
  networks,'' in \emph{Proceedings of the fortieth annual ACM symposium on
  Theory of computing}.\hskip 1em plus 0.5em minus 0.4em\relax ACM, 2008, pp.
  295--304.

\bibitem[Jackson and Rogers(2005)]{jackson2005economics}
M.~O. Jackson and B.~W. Rogers, ``The economics of small worlds,''
  \emph{Journal of the European Economic Association}, vol.~3, no. 2-3, pp.
  617--627, 2005.

\bibitem[Fricke and Lux(2012)]{fricke2012core}
D.~Fricke and T.~Lux, ``Core--periphery structure in the overnight money
  market: evidence from the e-mid trading platform,'' \emph{Computational
  Economics}, vol.~45, no.~3, pp. 359--395, 2012.

\bibitem[Jackson and Wolinsky(1996)]{jackson1996strategic}
M.~O. Jackson and A.~Wolinsky, ``A strategic model of social and economic
  networks,'' \emph{Journal of economic theory}, vol.~71, no.~1, pp. 44--74,
  1996.

\bibitem[Osborne and Rubinstein(1994)]{osborne1994course}
M.~J. Osborne and A.~Rubinstein, \emph{A course in game theory}.\hskip 1em plus
  0.5em minus 0.4em\relax MIT press, 1994.

\bibitem[Persitz(2010)]{persitz2010core}
D.~Persitz, ``Core-periphery r\&d collaboration networks,'' Working paper,
  Tech. Rep., 2010.

\bibitem[Johnson and Gilles(2003)]{johnson2003spatial}
C.~Johnson and R.~P. Gilles, ``Spatial social networks,'' in \emph{Networks and
  Groups}.\hskip 1em plus 0.5em minus 0.4em\relax Springer, 2003, pp. 51--77.

\bibitem[Bala and Goyal(2000)]{bala2000noncooperative}
V.~Bala and S.~Goyal, ``A noncooperative model of network formation,''
  \emph{Econometrica}, vol.~68, no.~5, pp. 1181--1229, 2000.

\bibitem[Bloch and Jackson(2007)]{bloch2007formation}
F.~Bloch and M.~O. Jackson, ``The formation of networks with transfers among
  players,'' \emph{Journal of Economic Theory}, vol. 133, no.~1, pp. 83--110,
  2007.

\bibitem[Papadimitriou(2001)]{papadimitriou2001algorithms}
C.~Papadimitriou, ``Algorithms, games, and the internet,'' in \emph{Proceedings
  of the thirty-third annual ACM symposium on Theory of computing}.\hskip 1em
  plus 0.5em minus 0.4em\relax ACM, 2001, pp. 749--753.

\bibitem[Galeotti et~al.(2006)Galeotti, Goyal, and
  Kamphorst]{galeotti2006network}
A.~Galeotti, S.~Goyal, and J.~Kamphorst, ``Network formation with heterogeneous
  players,'' \emph{Games and Economic Behavior}, vol.~54, no.~2, pp. 353--372,
  2006.

\bibitem[Vandenbossche and Demuynck(2013)]{vandenbossche2013network}
J.~Vandenbossche and T.~Demuynck, ``Network formation with heterogeneous agents
  and absolute friction,'' \emph{Computational Economics}, vol.~42, no.~1, pp.
  23--45, 2013.

\bibitem[Heydari et~al.(2015)Heydari, Mosleh, and Dalili]{heydari2015efficient}
B.~Heydari, M.~Mosleh, and K.~Dalili, ``Efficient network structures with
  separable heterogeneous connection costs,'' \emph{Economics Letters}, vol.
  134, pp. 82--85, 2015.

\bibitem[Heydari and Dalili(2015)]{heydari2015emergence}
B.~Heydari and K.~Dalili, ``Emergence of modularity in system of systems:
  Complex networks in heterogeneous environments,'' \emph{Systems Journal,
  IEEE}, vol.~9, no.~1, pp. 223--231, 2015.

\bibitem[Newman and Girvan(2004)]{newman2004finding}
M.~E. Newman and M.~Girvan, ``Finding and evaluating community structure in
  networks,'' \emph{Physical review E}, vol.~69, no.~2, p. 026113, 2004.

\bibitem[Kruskal(1956)]{kruskal1956shortest}
J.~B. Kruskal, ``On the shortest spanning subtree of a graph and the traveling
  salesman problem,'' \emph{Proceedings of the American Mathematical society},
  vol.~7, no.~1, pp. 48--50, 1956.

\bibitem[Graham and Hell(1985)]{graham1985history}
R.~L. Graham and P.~Hell, ``On the history of the minimum spanning tree
  problem,'' \emph{Annals of the History of Computing}, vol.~7, no.~1, pp.
  43--57, 1985.

\bibitem[Chevaleyre et~al.(2006)Chevaleyre, Dunne, Endriss, Lang, Lemaitre,
  Maudet, Padget, Phelps, Rodriguez-Aguilar, and Sousa]{chevaleyre2006issues}
Y.~Chevaleyre, P.~E. Dunne, U.~Endriss, J.~Lang, M.~Lemaitre, N.~Maudet,
  J.~Padget, S.~Phelps, J.~A. Rodriguez-Aguilar, and P.~Sousa, ``Issues in
  multiagent resource allocation,'' \emph{Informatica}, vol.~30, no.~1, 2006.

\bibitem[M{\"o}hlmeier et~al.(2013)M{\"o}hlmeier, Rusinowska, and
  Tanimura]{mohlmeier2013degree}
P.~M{\"o}hlmeier, A.~Rusinowska, and E.~Tanimura, ``A degree-distance-based
  connections model with negative and positive externalities,'' \emph{Bielefeld
  Working Papers in Economics and Management}, 2013.

\end{thebibliography}
\end{document}